\documentclass[conference]{IEEEtran}
\IEEEoverridecommandlockouts
\usepackage{cite}
\usepackage{amsmath,amssymb,amsfonts}
\usepackage{algorithmic}
\usepackage{graphicx}
\usepackage{textcomp}
\usepackage{xcolor}
\usepackage{hyperref}
\def\BibTeX{{\rm B\kern-.05em{\sc i\kern-.025em b}\kern-.08em
    T\kern-.1667em\lower.7ex\hbox{E}\kern-.125emX}}
\begin{document}

\title{C3S Micro-architectural Enhancement: Spike Encoder Block and Relaxing Gamma Clock (Asynchronous)}

\makeatletter
\newcommand{\linebreakand}{%
  \end{@IEEEauthorhalign}
  \hfill\mbox{}\par
  \mbox{}\hfill\begin{@IEEEauthorhalign}
}
\makeatother

\author{\IEEEauthorblockN{Alok Anand}
\IEEEauthorblockA{\textit{Electrical \& Computer Engineering} \\
\textit{Carnegie Mellon University}\\
Pittsburgh, PA, U.S.A. 15213 \\
\href{mailto:aloka@andrew.cmu.edu}{aloka@andrew.cmu.edu}}\\
\and
\IEEEauthorblockN{Ivan Khokhlov}
\IEEEauthorblockA{\textit{Electrical \& Computer Engineering} \\
\textit{Carnegie Mellon University}\\
Pittsburgh, PA, U.S.A. 15213 \\
\href{mailto:ikhokhlo@andrew.cmu.edu}{ikhokhlo@andrew.cmu.edu}}\\
\and
\IEEEauthorblockN{Abhishek Anand}
\IEEEauthorblockA{\textit{Electrical \& Computer Engineering} \\
\textit{Carnegie Mellon University}\\
Pittsburgh, PA, U.S.A. 15213 \\
\href{mailto:aanand3@andrew.cmu.edu}{aanand3@andrew.cmu.edu}}\\
\linebreakand

}

\maketitle

\begin{abstract}
  The field of neuromorphic computing is rapidly evolving. As both biological accuracy and practical implementations are explored, existing architectures are modified and improved for both purposes. The Temporal Neural Network(TNN) style of architecture is a good basis for approximating biological neurons due to its use of timed pulses to encode data and a voltage-threshold-like system. Using the Temporal Neural Network cortical column C3S architecture design as a basis, this project seeks to augment the network’s design. This project takes note of two ideas and presents their designs with the goal of improving existing cortical column architecture. One need in this field is for an encoder that could convert between common digital formats and timed neuronal spikes, as biologically accurate networks are temporal in nature. To this end, this project presents an encoder to translate between binary encoded values and timed spikes to be processed by the neural network. Another need is for the reduction of wasted processing time to idleness, caused by lengthy Gamma cycle processing bursts. To this end, this project presents a relaxation of Gamma cycles to allow for them to end arbitrarily early once the network has determined an output response. With the goal of contributing to the betterment of the field of neuromorphic computer architecture, designs for both a binary-to-spike encoder, as well as a Gamma cycle controller, are presented and evaluated for optimal design parameters, with overall system gain and performance.
\end{abstract}

\vspace{1em}

\section{Introduction}
Though research in the field of neural networks dates back decades, attempts to create a truly biologically accurate network architecture are relatively novel. This field of architecture research, known as neuromorphic computing, is fresh and rapidly evolving. In recent years companies have sprung up that focus on producing neural processors for various purposes such as accelerating learning tasks. The field has great potential in applications from task acceleration to the simulation of animal-like neural processes. In the great race for improving artificial intelligence, neuromorphic computer architecture is potentially a holy grail for the realization of that goal; following the idea that “what can learn the world better than animals” – creatures that have evolved over millennia to learn and make sense of the world they live in. 
 
Neuromorphic architecture designs have been adjusted over the years, and the general consensus today is on the need for a TNN with Spike-Time Dependent Plasticity (STDP) for updates of the inter-neuronal synapse weights. The C3S architecture produced at Carnegie Mellon University \cite{9516717}, is one such architecture that has evolved much in recent years and is constantly being augmented. This project focuses on novel augmentations to this architecture for the purpose of not only improving the design but also furthering the field of neuromorphic architecture research.

\subsection{What is an Encoder?}
An encoder converts input data into a range of representations for computation. It provides a method to biological and physical data such as images into valuable data sources, capturing essential data markers and features. The encoder representation of input data into data bits 0, and 1 precisely maps the characteristics of data that match to application.
Designing an encoder needs determining and extracting key information. In this work, handling imaging need to read the image pixel data with each data point corresponding to the brightness amplitude that determines the corner of the image in a well-defined image. 
In a good encoder design, as suggested by Scott in \cite{article}, it is required to keep all the input data points intact during conversion with efficient mapping to encoded output data. Our work focuses on a positive-negative encoder that produces a positive and negative form of image pixel data based on a threshold with each iteration of input producing the same output result. With every iteration on the same data and maintaining consistency in the dimension of encoded output data bits to input data, for all of the data points.

\subsection{Gamma Cycle Oscillation}
Gamma Cycle corresponds to rhythmic synchronization in the frequency range of $30$ to $100$ Hz range as reported by Pascal et al. \cite{FRIES2007309}, for  temporal conversion of input data that arrives as sensory stimuli. Neural process sequential input arriving every gamma cycle is mapped to the column that converts it to temporal form. The amplitude of input stimuli to every column is recoded and converted to a temporal value as depicting the time of occurrence and correlation between neural units as proposed by Burwick \cite{Burwick2009GammaOA}. Columns comprising neurons generate spikes from within a column based on accumulating spike responses, with stronger input sources firing with an earlier response spike. Within a gamma cycle time window, columns capture excitatory input stimuli, to accumulate output spikes until all neurons in the column spike and once a column is spiked, the winner takes all, inhibiting other columns from spiking and the whole column network is inhibited until the next new gamma cycle start.

\section{Motivation}
\subsection{Encoder Design}
In this work we present the first micro-architectural implementation, and verification of the positive-negative (posneg) encoder system, already modeled and verified with C3S architecture work by Harideep et al \cite{9516717}. The basic unit of posneg encoder design includes a data flow unit to schedule images for processing and the comparator unit capable of generating 1-bit (either 0 or 1) output based on a threshold value for the input.

\subsection{Asynchronous Gamma-cycle oscillation}
In a defined time period, synchronous gamma cycle significance, as mentioned in biological overview by Smith \cite{Smith2017}, is drawn as the greater the input stimuli amplitude, the higher spike frequency with greater input amplitude leads to faster/sooner spikes and performs STDP weight learning for other column neurons accordingly. Moving towards asynchronous gamma cycle to end gamma cycle as soon as a column spiked and performing weight update to start new gamma cycle. This design would reduce training time and weight update for learning larger datasets and drawing inferences.

\section{Main Idea}
\subsection{Positive-Negative encoder}
In a positive-negative encoder, input data is compared against a threshold value and assigned with a corresponding data value of 0 and 1 representing every input data point amplitude that highlights the image corner. With positive encoding comparing data points with a threshold value of $127$, and assigning value $1$ with pixel data amplitude greater than the threshold and value $0$ for pixel data less than the threshold.
In negative encoding scheme data points when compared with a threshold $127$ and with a value greater than a threshold are assigned with value $0$ and data points with a value less than the threshold are assigned with $1$.
This encoding representation seems to be easy to scale based on image pixel and flexible which makes it easy to integrate with synapses column input (neurons) in C3S micro-architectural design. 

\subsection{Asynchronous Gamma cycle oscillations}
With increases in excitation input stimuli amplitude, the threshold is reached sooner than the completion of the gamma cycle period, here 16 clock cycles, and a column fires with output spike, inhibiting spikes from other columns and performing weight update in spike time-dependent plasticity until the next gamma cycle. All this gamma cycle to column generating spikes, inhibiting other columns spikes with winner takes all and self-terminating of gamma cycle to start new cycle takes place synchronously with fixed gamma cycle duration. 
In column neurons that don't spike within the gamma cycle time frame, in that case, columns will not be able to fire until the next gamma cycle.
In the synchronous gamma cycle, the controlling trigger of the output spike, when the threshold is reached earlier before the gamma spike inhibits the column from responding to other input spikes and STDP learning till the next fresh gamma cycle starts. As a result of which synchronous gamma cycle-based STDP learning and weight update tend to always incur the same latency and energy costs even in some cases where the threshold is reached earlier.

\section{Architectural Implementation}
\subsection{Positive-Negative Encoder}
Processing real-time images and videos involves a large number of amount of images
to be buffered and scheduled with low latency to be relevant in real-time. This work addresses, the design required for a large number of image data that is to be fetched, scheduled, and processed.
Along with the encoding required for C3S architecture for processing Images or MNIST data, the factors that constrain the encoder design are the time to process an image for real-time edge applications, with energy and area requirements to make it feasible for neuromorphic application. 

The data flow unit as shown in \textit{Fig.\ref{fig:encoder_dataFlow}}, buffers the entire pixel of an image, so in every unit clock cycle, sampling a few pixels of the buffered image and sending it to the comparator unit for encoding. Sample pixel count is determined by the number of available comparator units.

\begin{figure}[h!]
    \begin{center}    \includegraphics[width=0.5\textwidth]{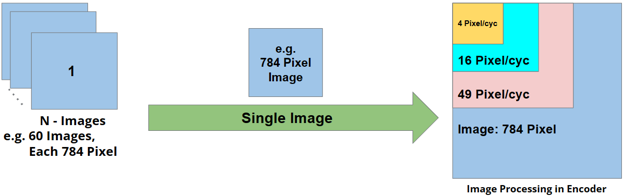}
    \end{center}
    \caption{Encoder Data Flow}    \label{fig:encoder_dataFlow}
\end{figure}

In the block diagram, as in \textit{Fig.\ref{fig:encoder_model}} of encoder implementation, input comprises of image dataset of configurable image pixel size that upon reading is sampled in smaller pixel size to the comparator. This process continues until the entire image is sampled and encoded to produce a negative and positive encoded image as output.

\begin{figure}[h!]
    \begin{center}    \includegraphics[width=0.5\textwidth]{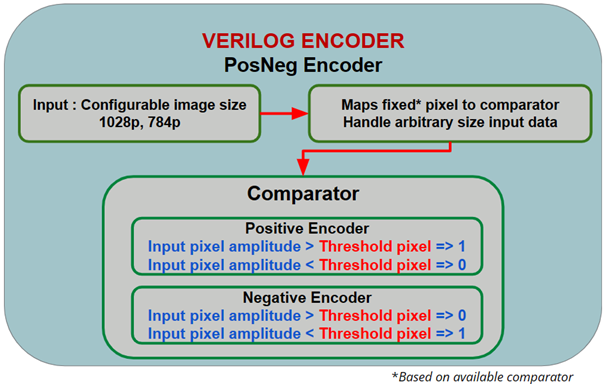}
    \end{center}
    \caption{Encoder Model}    \label{fig:encoder_model}
\end{figure}

As shown in figure \textit{Fig.\ref{fig:encoder_tb_model}}, a block diagram of the test-bench for the pos-neg encoder unit that includes a Python script \cite{3}, to process MNIST files and system verilog modules to feed MNIST images to the encoder module. Test-bench starts with a python script, takes MNIST file $ubyte$ \cite{4}, as input to read each image, and writes image pixel in a text $.txt$ file. Each line of the text file holds the entire pixel of an Image. The script has configurable values for image count, the number of pixels in an image lead to image size.
In generated $.txt$ file from the Python script is input to SystemVerilog test-bed that is parameterized with a number of images in a text $.txt$, file, image pixel size, and pixel threshold (pos-neg encoding) values. All the images in the text file are read at once to store in a linear 2D matrix variable. Images stored in the matrix have sampled an image in each unit clock cycle to send it to the encoder unit. Once an image is sent to the encoder unit, the next image waits for further clock cycles, while observing the output signal from the encoder unit. This output signal from the encoder unit shows the processing of the image by the encoder unit and marks it as a signal to test-bed, so as to send the next image. 
Each encoded output of the image from the encoder is written in the linear 2D matrix variable which is further written to memory or in this case to the output text $.txt$ file.

\begin{figure}[h!]
    \begin{center}    \includegraphics[width=0.5\textwidth]{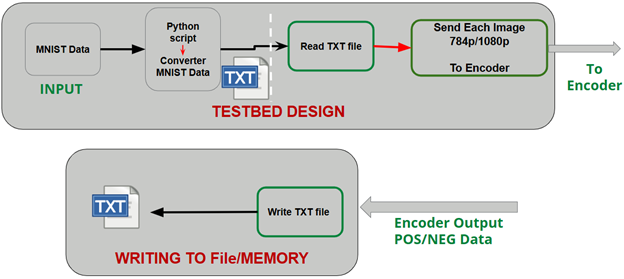}
    \end{center}
    \caption{Encoder test bench model}    \label{fig:encoder_tb_model}
\end{figure}

\subsection{Relaxed Gamma Cycle Control}

Relaxed Gamma Cycle Control was accomplished through the creation of two SystemVerilog Modules: grst\_generator and grst\_controller. 

The generator module flow depicted in \textit{Fig. \ref{fig:generator}} was essentially an up counter, which upon reaching the value of the gamma cycle period, would generate gamma reset, $grst$ signal and reset its counter, thus starting a new gamma cycle. Additionally, if it were to receive a control signal from the controller module, it would generate the $grst$ signal early and reset its counter, to shorten a gamma cycle through a control signal.

\begin{figure}[h!]
    \begin{center}    \includegraphics[width=0.5\textwidth]{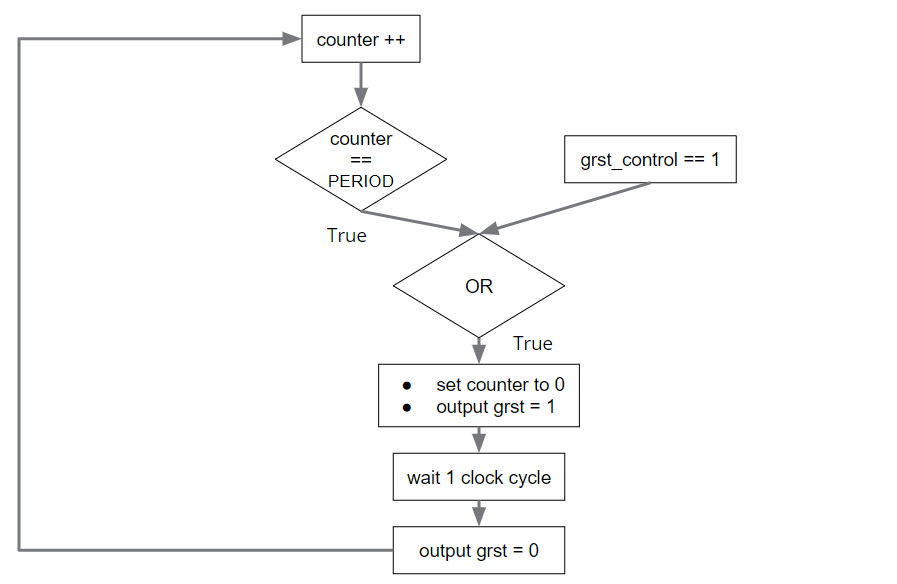}
    \end{center}
    \caption{Flowchart depicting the logical flow of the grst\_generator module}    \label{fig:generator}
\end{figure}

The controller module in \textit{Fig.\ref{fig:spike_checker}} and \textit{Fig.\ref{fig:controller}} is designed to track output spikes from columns in the network, and to signal to the grst generator that it should start a new gamma cycle if all columns have generated an output. 
\begin{figure}[h!]
    \begin{center}    \includegraphics[width=0.5\textwidth]{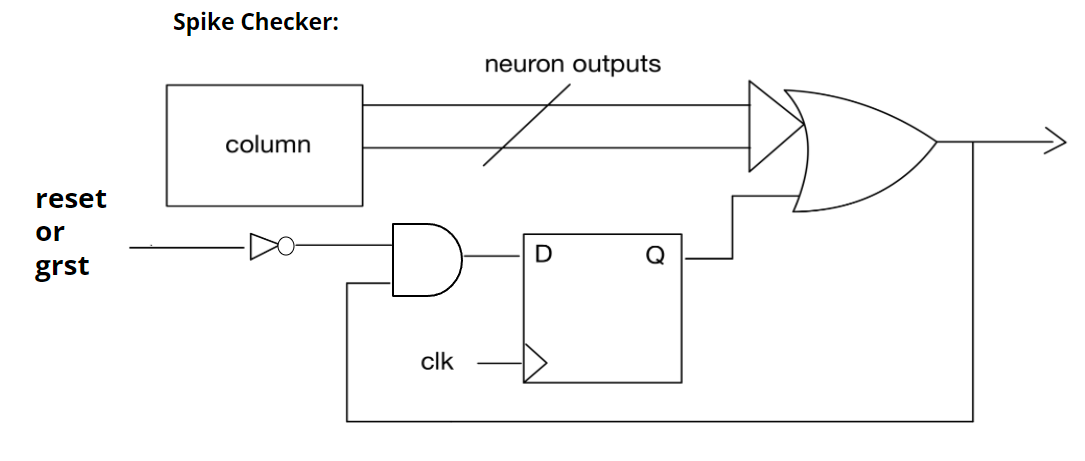}
    \end{center}
    \caption{Diagram showing the OR structure surrounding every column's neuron outputs for the purpose of storing if that column has spiked yet. Termed "Spike Checker" in later figures. }    \label{fig:spike_checker}
\end{figure}
It essentially consists of large OR gates over each column, taking in the outputs and generating a 1 should any neuron spike \textit{Fig.\ref{fig:spike_checker}}. Additionally, each column would have a flip-flop assigned to it, that would store 1 as output of the OR gate. Thus, if an OR gate ever produced a 1, it would always produce a 1 until the next Gamma cycle. Finally, all OR gate outputs would be $ANDed$ together to produce the grst control output \textit{Fig.\ref{fig:controller}}. Thus, if there has been at least one spike in each column, the controller tells the generator to start a new gamma cycle.

\begin{figure}[h!]
    \begin{center}    \includegraphics[width=0.5\textwidth]{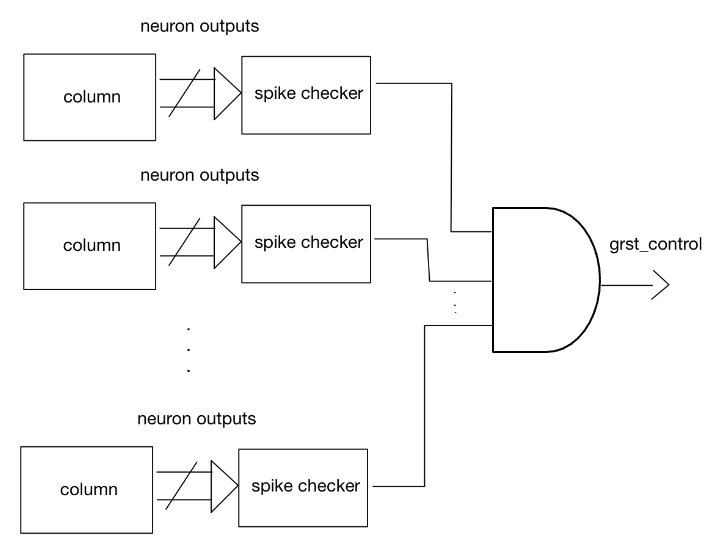}
    \end{center}
    \caption{Diagram showing the overarching AND structure surrounding all OR outputs connected to all columns leading to the production of the grst\_control signal of the grst\_controller module. }    \label{fig:controller}
\end{figure}

When integrated into a multi-layer TNN, the controller should look at only the final layer’s outputs \textit{Fig. \ref{fig:multi_layer}}. Thus when the entire network has produced all outputs, a new Gamma cycle can begin, saving processing time in the process.

\begin{figure}[h!]
    \begin{center}    \includegraphics[width=0.5\textwidth]{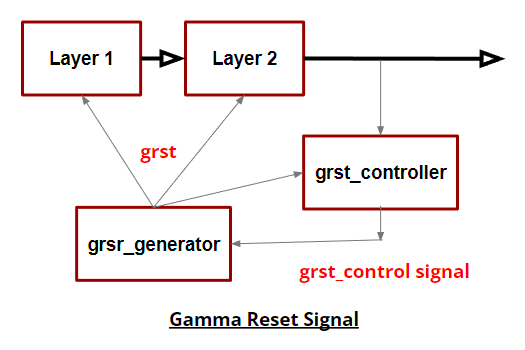}
    \end{center}
    \caption{Diagram showing the integration of grst generator and controller modules into a multi-layer TNN.}    \label{fig:multi_layer}
\end{figure}

\section{Methodology}
In the micro-architectural implementation of the modules, our work used Synopsys design compiler for Verilog logic design and simulation. The simulation includes functional verification of individual modules with a test bench that includes optimization for latency and area costs.
Hardware synthesis of functional modules and design evaluation are performed with a design compiler for power, area, and timing values under different parameters.
For the logic synthesis of design, this work used a $45 nm$ technology node with Nangate standard cell library with $1$GHz clock frequency for PPA analysis.

\subsection{Encoder}
In encoder design, hardware synthesis includes the evaluation of the comparator to sample the smallest pixel covered per clock cycle as a part of the entire image of 784 pixels, ranging from the smaller pixel of 2x2, 4x4 to the larger pixel of 28x28.

\subsection{Gamma Cycle Control Architecture}
In the asynchronous gamma cycle, our work includes hardware synthesis of gamma generator and gamma controller functional modules. The gamma generator module is parameterized for a period of 16 clock cycles and analyzed for area, power, and path delay from the design compiler.

The Gamma cycle controller and generator modules were designed using SystemVerilog and evaluated using both Synopsys VCS simulation and Design Compiler. 

The design followed a two-module structure, where one module would produce the gamma reset signals on a set period, and the other would tell the first module to produce a Gamma reset signal early should all columns produce an output spike.

In this work, VCS is used for functional verification, conducting several tests to prove the proper workings of the generator, controller, and both as a system.

Using Design compiler, to determine the physical properties of the designs and to see how these properties scale with adjustments to column and neuron-per-column quantities.

\subsection{Justifying Relaxed Gamma Cycles}

To justify the potential benefits of reducing gamma cycles, tests were performed to demonstrate how RNL(Ramp-no-leak) layers, once trained, could stabilize at some point less than the maximum gamma cycle period.

To accomplish this, the threshold was raised from 400 to 4,000, as at any value under around 700, RNL neurons would always produce output spikes at time 0 due to the fact that with a PosNeg encoding, there are always approximately 700 time 0s being passed in as inputs to the columns. Additionally, two other encodings, Linear and Logarithmic, were tested for the purpose of potentially seeing more of a "training curve" where layer outputs slowly learn to spike faster and faster. 

All encoders tested for spike-time occurrences were of a Pos-Neg nature. That is to say, for Linear and Logarithmic encoding, there was one encoding done over the values, and a second done as a negative of that first linear encoding, e.g.[0 1 2 3 4 5] become [$\infty$ 15 14 13 12 11] in the negative. The values were first negated and then again encoded using linear or logarithmic. The exact formula used is expressed as:

\begin{equation}
\text{encode}(|\text{tensor} - \text{maxt}|)
\end{equation}

where encode would be the encoding algorithm, and maxt would be the maximum possible value of the tensor i.e.1.0.

\section{Experimental Design}
\subsection{Encoder}
\subsubsection{posnegEncoder}
In the comparator module shown in \textit{Fig.\ref{fig:comparator}} takes an 8-bit pixel of an image as $inVal$ as input to both. The module has a positive and a negative comparator. This input pixel is compared with the threshold pixel to generate 1-bit output. In the positive comparator, if the input pixel is greater than the threshold pixel return 1 as output and 0 if less than the threshold pixel. Similarly for the negative comparator, if the input pixel is greater than the threshold pixel value returns 0 as output and 1 if less than the threshold pixel.
\begin{figure}[h!]
    \begin{center}    \includegraphics[width=0.5\textwidth]{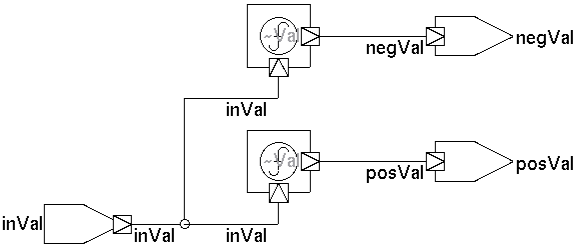}
    \end{center}
    \caption{posnegEncoder module}    \label{fig:comparator}
\end{figure}

\subsubsection{comparator\_parallel\_ units}
As shown in \textit{Fig.\ref{fig:comparator_count49}}, upper hierarchical Verilog implementation of encoder body block is implemented as $49$ parallel comparator block encoding 7x7 pixel of the image at each clock cycle by cycle as $posout$ and $negout$.

\begin{figure}[h!]
    \begin{center}    \includegraphics[width=0.5\textwidth]{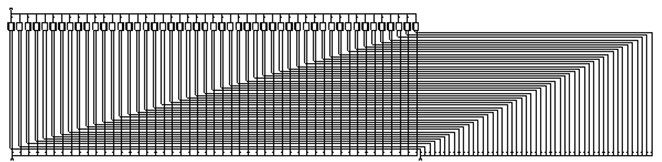}
    \end{center}
    \caption{Positive and Negative encoder unit with 49 parallel comparators}    \label{fig:comparator_count49}
\end{figure}

\subsubsection{data\_encoder} 
Data encoder is shown in \textit{Fig.\ref{fig:decoderUnit}} that image in port $image\_in$ and $clk$ signal serves as the unit clock. This block samples the configurable sample size of the image pixel at every clock cycle and provides as input $in$ port to the posnegEncoder block that generates positive and negative encoded image compared with threshold value as an output $posout$ and $negout$ to $Data\_encoder$, which accumulates this encoded result cycle by cycle until the entire image is processed, to generate the final encoded output image at port $image\_posout$ with the positive encoded result and $image\_negout$ with the negative encoded result.

\begin{figure}[h!]
    \begin{center}    \includegraphics[width=0.5\textwidth]{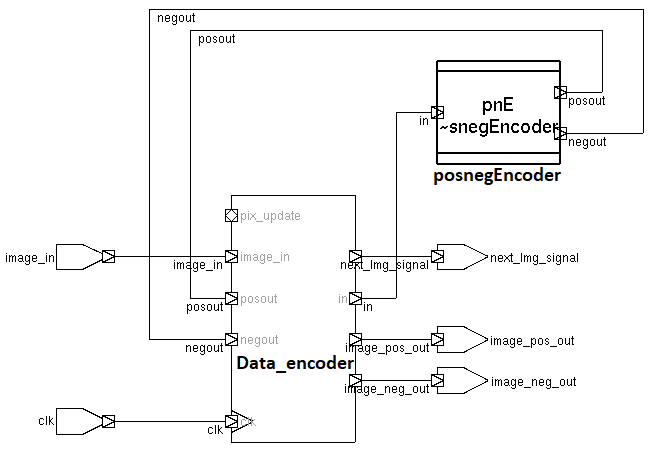}
    \end{center}
    \caption{data\_encoder module}    \label{fig:decoderUnit}
\end{figure}

\subsubsection{decoder\_tb}
The figure \textit{Fig. \ref{fig:decoder_tb}} shows the hierarchical implementation of the encoder along with the testbed, with $Pixel\_linear$ block reading and loading image via port $imagePixel\_linear$ and passing image within a clock cycle to $data Encoder$ block that samples a configurable smaller pixel block image each cycle defined by $aclk$ to generate $image\_posout$ and $image\_negout$. Block $image\_reached$ uses this information to conclude that image processing is performed with $imageReached$ output signal to $main\_block$ to start loading other images and increasing image count $imgCount$.

\begin{figure}[h!]
    \begin{center}    \includegraphics[width=0.5\textwidth]{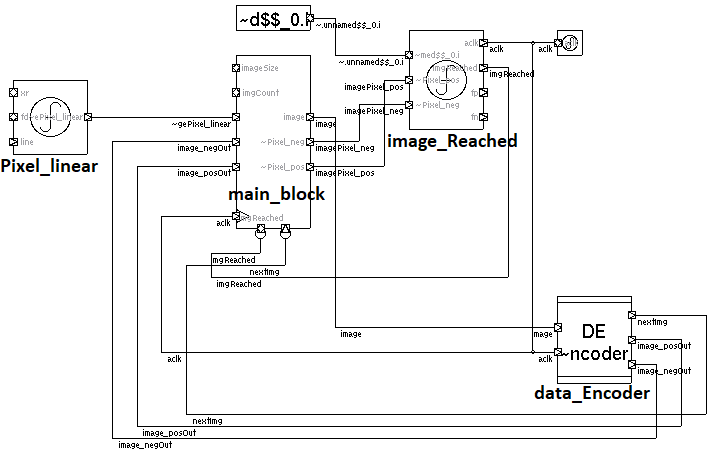}
    \end{center}
    \caption{Encoder\_tb module}    \label{fig:decoder_tb}
\end{figure}

\subsection{Gamma Cycles}
Three main functional system tests were conducted for the Gamma cycle modules:
\begin{enumerate}
    \item grst generating when all neurons spike simultaneously
    \item grst generating when neurons spike one by one at different times
    \item grst generating on the normal interval, no column output spikes
\end{enumerate}
For the justification of potential benefits, reference \cite{9516717} is modified to accommodate the two new encodings and evaluated. Two tests were performed for each encoder, one at the default threshold of 400, and one at the elevated threshold of 4000. Values extracted from these tests were: the quantities of spikes that occurred at X time; and the purity of the encoder. 

\subsubsection{grst\_generator}
The figure \textit{Fig.\ref{fig:grst_generator}} below, shows the RTL design of gamma reset generator $grst$ that is basically a counter module with parameterizable period parameter. In the generator of grst pulses counter overflow $ORed$ with a control input from grst controller module output results in a grst pulse.

\begin{figure}[h!]
    \begin{center}    \includegraphics[width=0.5\textwidth]{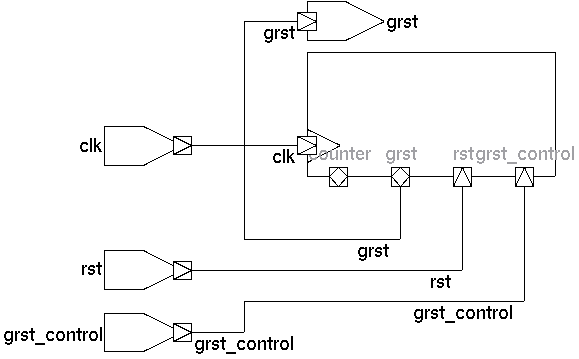}
    \end{center}
    \caption{grst\_generator module}    \label{fig:grst_generator}
\end{figure}

\subsubsection{grst\_controller}
In \textit{Fig.\ref{fig:grst_controller_logic}} that depicts RTL implementation that listens to outputs of all columns, wait for at least one output pulse from each column, and sends a control signal to grst generator to output as a pulse on the next positive edge of the clock, with number of columns and pulse period are customizable

\begin{figure}[h!]
    \begin{center}    \includegraphics[width=0.5\textwidth]{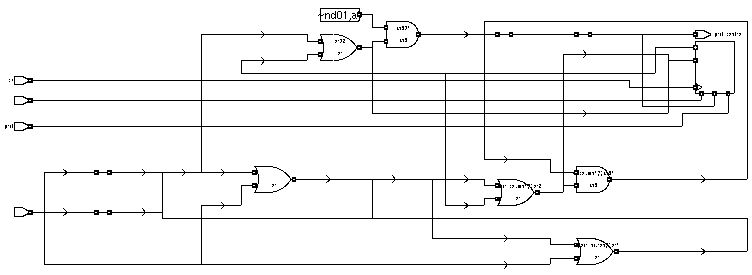}
    \end{center}
    \caption{grst\_controller module}    \label{fig:grst_controller_logic}
\end{figure}

Power, area, and propagation delay values were computed using Synopsys Design Compiler. These were verified both through $monitor$ outputs and waveforms.

\subsection{STDP weight update}
In the \textit{Fig.\ref{fig:stdp_casegen_logic}} shows case generation logic for additional STDP weight update, when synapse in the column exceeds the threshold and generates a spike and ends gamma cycle earlier than gamma cycle period to perform weight update, for input condition specified in \textit{Table\ref{tab7}} that results in relaxing gamma cycle to asynchronous.

\begin{figure}[h!]
    \begin{center}    \includegraphics[width=0.5\textwidth]{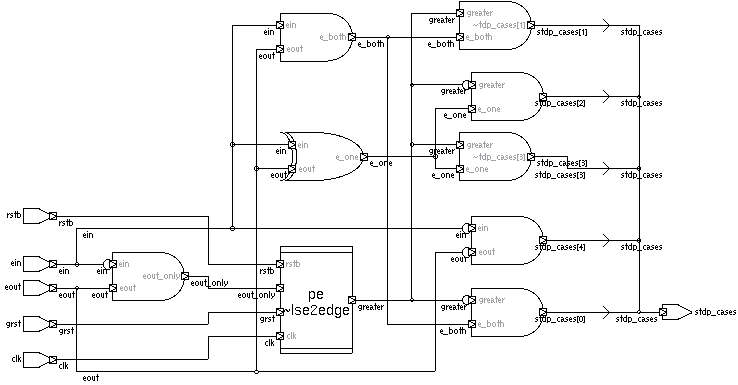}
    \end{center}
    \caption{stdp\_casegen module}    \label{fig:stdp_casegen_logic}
\end{figure}

STDP increment and decrement logic in \textit{Fig.\ref{fig:stdp_incdec_logic}} implement weight update rule incrementing weight with $0.5u$ during each gamma cycle, for the cases when there is no input signal and output spike.

\begin{table}[htbp]
\caption{Additional STDP weight update: no input and no output case}
\begin{center}
\begin{tabular}{|c|c|}

\hline
\textbf{Input conditions} & \textbf{Weight Update rule} \\

\hline
\text{x(t) = $\infty$ ; z(t) = $\infty$}  & \text{+0.5u} \\
\hline

\end{tabular}
\label{tab7}
\end{center}
\end{table}

\begin{figure}[h!]
    \begin{center}    \includegraphics[width=0.5\textwidth]{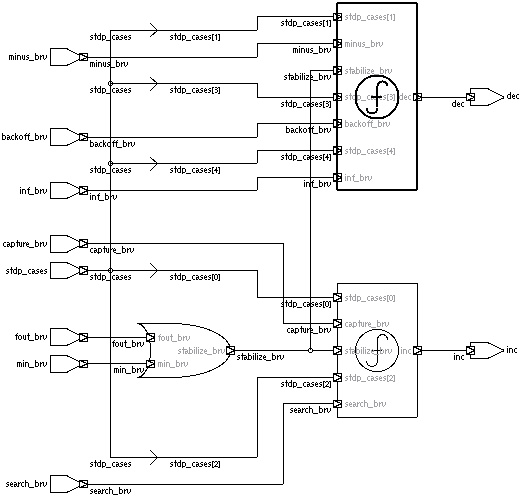}
    \end{center}
    \caption{stdp\_incdec module}    \label{fig:stdp_incdec_logic}
\end{figure}

\textit{Table.\ref{tab1}} reports respective PPA (Power-Performance-Area) metrics of STDP case generation logic and presents results from $45nm$ node at 1 GHz clock with performance measured in terms of computation time, critical path delay, and area. With an area as total cell area in sq-nm and total power accounts for the sum of dynamic and leakage power in $uW$.

\begin{table}[htbp]
\caption{PPA data for stdp case generation logic}
\begin{center}
\begin{tabular}{|c|c|c|}

\hline
\textbf{Area (sq nm)} & \textbf{Power (uW)} & \textbf{Critical Path Delay (ns)} \\

\hline
\text{8.778} & \text{4.7499} & \text{2.16} \\
\hline

\end{tabular}
\label{tab1}
\end{center}
\end{table}

\textit{Table.\ref{tab2}} reports PPA data for weight increment and decrement in STDP, uses $45nm$ technology and 1 GHz clock, with hardware implementation measured in total cell area sq-nm, critical data path in $ns$ and power in $uW$.

\begin{table}[htbp]
\caption{PPA data for weight incdec logic}
\begin{center}
\begin{tabular}{|c|c|c|}

\hline
\textbf{Area (sq nm)} & \textbf{Power (uW)} & \textbf{Critical Path Delay (ns)} \\

\hline
\text{7.182} & \text{2.9621} & \text{0.14} \\
\hline

\end{tabular}
\label{tab2}
\end{center}
\end{table}

\section{Experimental Results}
\subsection{Encoder Analysis}

\subsubsection{Functional Verification}
\hfill

To perform the functional verification of the encoder design, we used the MNIST handwritten digit image database.
Python script is used to perform initial processing on the MNIST file to generate a text file with an image pixel value as shown in \textit{Fig.\ref{fig:encoder_input_image}}. Each line of the text file holds the entire pixel of an image.

\begin{figure}[h!]
    \begin{center}    \includegraphics[width=0.5\textwidth]{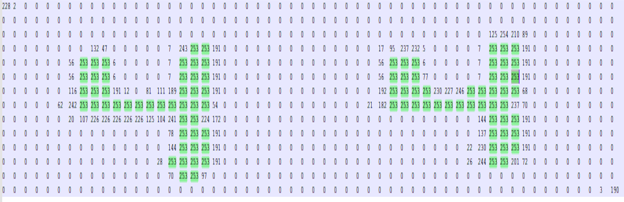}
    \end{center}
    \caption{Encoder input MNIST image representing numeric value 4 in image pixel value.}    \label{fig:encoder_input_image}
\end{figure}

The encoder test bed takes a text file as input which further reads each image from the file at a unit clock cycle and sends it to the encoder. Testbed waits to send the next image until the image send to the encoder completes the processing. The pos-neg encoded output from the encoder is written to two output files with one holding a positive encoded value as shown in \textit{Fig.\ref{fig:encoder_output_postive_image}} and the other holding a negative encoded value as shown in \textit{Fig.\ref{fig:encoder_output_negetive_image}}. The encoded output result confirms the functional correctness of the encoder's design.

\begin{figure}[h!]
    \begin{center}    \includegraphics[width=0.5\textwidth]{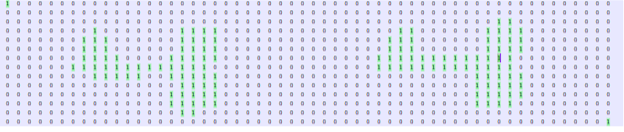}
    \end{center}
    \caption{Encoder positive output with pixel threshold of 127 for processing MNIST image representing numeric value 4 in image pixel value.}    \label{fig:encoder_output_postive_image}
\end{figure}

\begin{figure}[h!]
    \begin{center}    \includegraphics[width=0.5\textwidth]{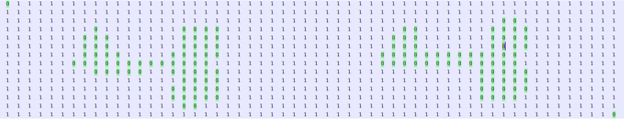}
    \end{center}
    \caption{Encoder negative output with pixel threshold of 127 for processing MNIST image representing numeric value 4 in image pixel value.}    \label{fig:encoder_output_negetive_image}
\end{figure}

\subsubsection{Design Costs}
\hfill

To assess the frequency dependence of the encoder design, we evaluated the design by varying clock frequencies ranging from 100MHz to 100GHz for an input image of 784 pixels and obtained PPA analysis data. Encoder hardware kept constant with 49 pos-neg comparator units with a constant area of 65.17 $sq-nm$ and a critical data path of 0.04 $ns$. Constant hardware leads to the dynamic power of 26.769 $uW$ and leakage power of 1.7538 $uW$. Until the unit clock cycle is larger than the critical data path of 0.04 $ns$, dynamic power remains constant but leakage energy is spent for the entire duration clock cycle,  so varying leakage energy decreased by increasing clock frequency due to reduced clock cycle time.  Leakage energy is comparatively higher than dynamic and dominates the total energy consumption by varying frequency increased $10^6$ times from 100KHz to 100GHz, so total energy decreases by 0.0002x  as in \textit{Fig.\ref{fig:encoderResult_varying_freq_energy}}. 
Image processing time depends inversely on clock cycle time, and decreases as frequency increases as in \textit{Fig.\ref{fig:encoderResult_varying_freq_processingtime}}.

\begin{figure}[h!]
    \begin{center}    \includegraphics[width=0.5\textwidth]{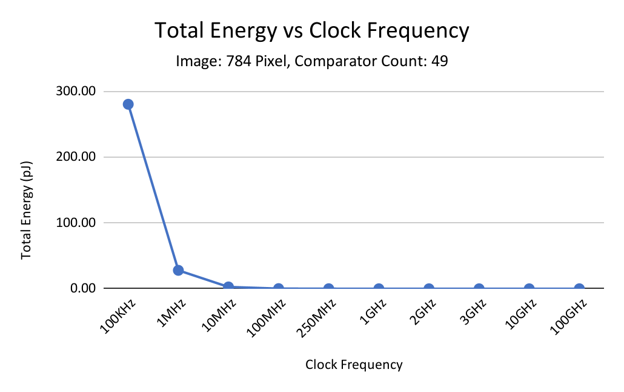}
    \end{center}
    \caption{Waveform analysis of total energy variation with respect to the operating clock frequency for processing 784-pixel image with 49 comparator count.}    \label{fig:encoderResult_varying_freq_energy}
\end{figure}

\begin{figure}[h!]
    \begin{center}    \includegraphics[width=0.5\textwidth]{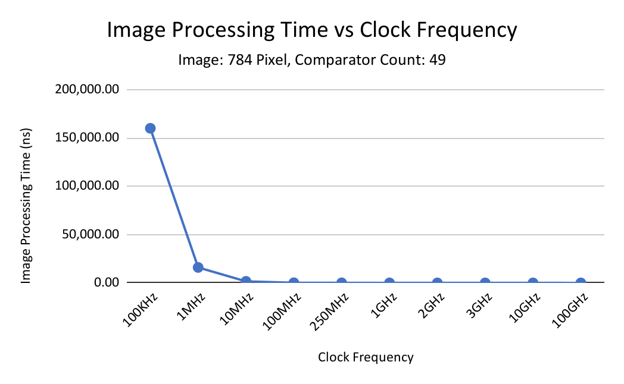}
    \end{center}
    \caption{Waveform analysis of processing time variation with respect to  the operating clock frequency for processing 784 pixel image with 49 comparator count.}    \label{fig:encoderResult_varying_freq_processingtime}
\end{figure}

To access the optimized comparator count, we evaluated the design to process 784 pixels at 1 GHz clock frequency, plotting power, energy, area, image processing time, and EDP (energy-delay product), for varying comparator counts (image pixel per cycle). 
The single comparator unit comprises two comparators (one for positive encoding and the other one for negative encoding) accounting for a cell area of 1.33 $sq-nm$, 546.3058 $nW$ of dynamic power with 35.7914 $nW$, and a data arrival time of 0.04 $ns$.

\begin{figure}[h!]
    \begin{center}    \includegraphics[width=0.5\textwidth]{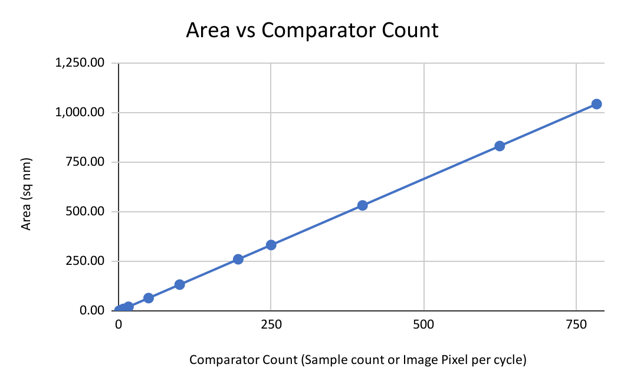}
    \end{center}
    \caption{Waveform analysis of encoder area with respect to the number of comparator.}    \label{fig:encoderResult_varingCmp_area}
\end{figure}

The plot in  \textit{Fig.\ref{fig:encoderResult_varingCmp_area}} shows comparator module scales linearly with the number of comparators. The area is the direct multiple of the comparator count with the unit comparator area.  For the comparator module with a 2x4 comparator to process a 784-pixel image size in 196 unit clock cycle at a clock frequency of 1 GHz is 4x times of single comparator unit area i.e 1.33 $sq-nm$ and similarly for a 2x784 comparator is 784x times of single comparator unit area i.e 1.33 $sq-nm$.

\begin{figure}[h!]
    \begin{center}    \includegraphics[width=0.5\textwidth]{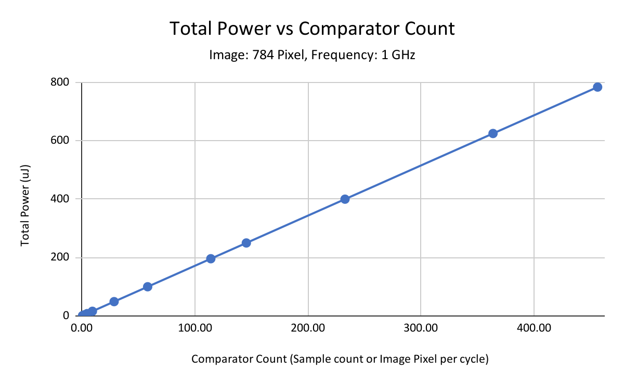}
    \end{center}
    \caption{Waveform analysis of power variation with respect to the number of comparator for processing 784 pixel image running at 1 GHz clock frequency.}    \label{fig:encoderResult_varingCmp_power}
\end{figure}

Power varies linearly as shown in \textit{Fig.\ref{fig:encoderResult_varingCmp_power}} with the comparator count and is the direct multiple of comparator count with unit comparator dynamic and leakage power of  546.3058 $nW$ and 35.7914 $nW$ respectively and carried the same for total power with unit power value 0.58 $uW$.

\begin{figure}[h!]
    \begin{center}    \includegraphics[width=0.5\textwidth]{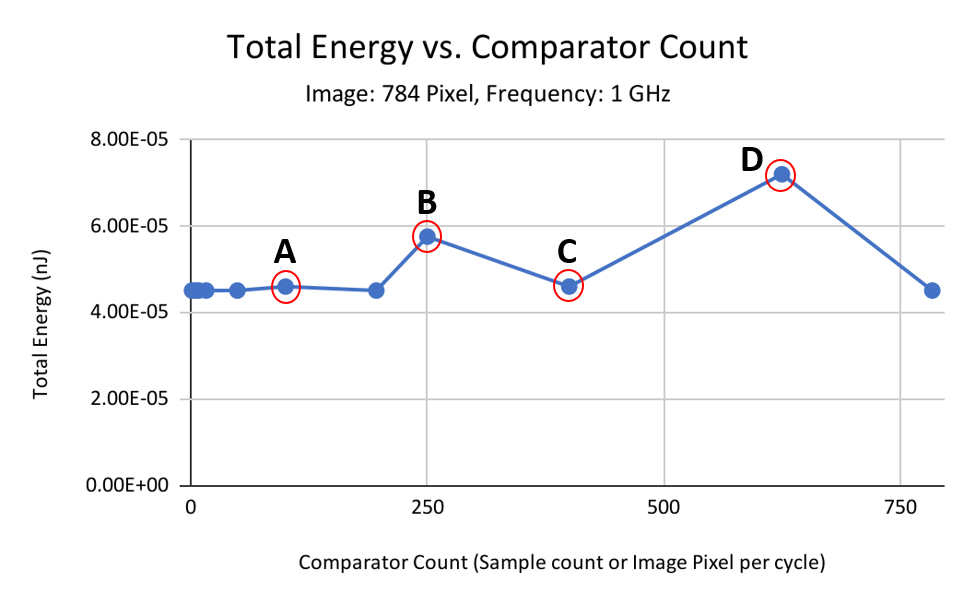}
    \end{center}
    \caption{Waveform analysis of energy variation with respect to the number of comparator for processing 784 pixel image running at 1 GHz clock frequency.}    \label{fig:encoderResult_varingCmp_energy}
\end{figure}

To study comparator count dependence on energy consumption, we evaluated the design while processing a 784-pixel image size running at 1 GHz. Dynamic, leakage and total energy required to process an image remained the same and are independent of comparator count, in case the comparator count is an exact divisor of the input image in the process. For the 784-pixel image size, the comparator count of 1, 2, 4, 8, 16, 49, 196, and 784 are exact divisors of 784, so the total energy is 4.52x$10^{-5}$ $nJ$ for all cases.
If the comparator count isn't exactly divided by pixel value as shown in \textit{Fig.\ref{fig:encoderResult_varingCmp_energy}} at points A, B C, and D where the comparator count is 100, 250, 400, and 625, the energy consumption is more due to the unnecessary use of comparators processing undefined data during last iteration of image processing. In the last iteration of image processing, all comparators are not used in the fixed comparator design.

\begin{figure}[h!]
    \begin{center}    \includegraphics[width=0.5\textwidth]{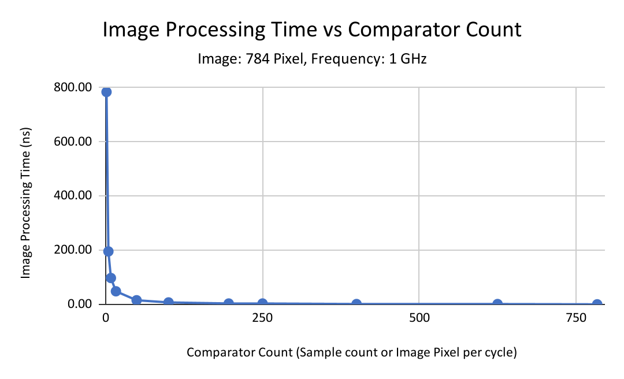}
    \end{center}
    \caption{Waveform analysis of processing time variation with respect to the number of comparator for processing 784 pixel image running at 1 GHz clock frequency.}    \label{fig:encoderResult_varingCmp_processingTime}
\end{figure}

The plot in \textit{Fig.\ref{fig:encoderResult_varingCmp_processingTime}} shows inverse dependence of processing time dependence with respect to the comparator count. As the comparator count increases processing time decreases since there are more comparators to process more image pixels in a cycle. When the clock frequency is 1 GHz, with one comparator unit, it takes 784 unit clock cycle i.e. 784$ns$, and 784 comparator unit takes 1 unit cycle i.e. 1$ns$ to process a 784 pixel image.

\begin{figure}[h!]
    \begin{center}    \includegraphics[width=0.5\textwidth]{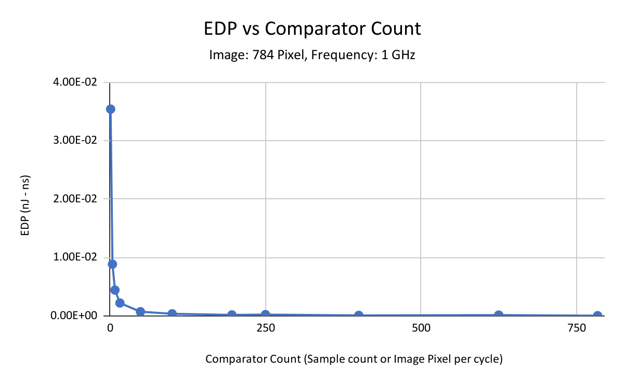}
    \end{center}
    \caption{Waveform analysis of EDP with respect to the number of comparator for processing 784 pixel image running at 1 GHz clock frequency.}    \label{fig:encoderResult_varingCmp_EDP}
\end{figure}

The plot in \textit{Fig.\ref{fig:encoderResult_varingCmp_EDP}} assess the energy-delay product to measure the energy efficiency and performance of the comparator. Data obtained from this plot depicts a pareto-optimal comparator dimension of less than 100 with increased EDP.

\subsubsection{Viability}
\hfill

To justify the usefulness and reflect the real-time behavior of our encoder design, we performed the evaluation with various-sized input images, the processing time of a certain number of images, and the image count processed in a given processing time. 

\begin{figure}[h!]
    \begin{center}    \includegraphics[width=0.5\textwidth]{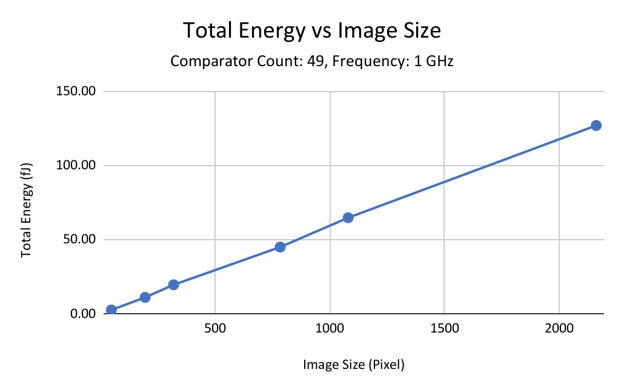}
    \end{center}
    \caption{Waveform analysis of total energy variation with respect to the varying input image size for 49 comparator count running at 1 GHz clock frequency.}    \label{fig:encoderResult_varying_imgSize_energy}
\end{figure}

While evaluating the design with respect to different input image sizes, we plotted the energy consumption and processing time as shown in \textit{Fig.\ref{fig:encoderResult_varying_imgSize_energy}} and \textit{Fig.\ref{fig:encoderResult_varying_imgSize_processingTime}} respectively. The design has a constant comparator count of 49 running at 1 GHz clock frequency. Both energy and processing time varies linearly and increases as Image size increases.

As shown in \textit{Fig.\ref{fig:encoderResult_varying_imgSize_energy}}, when the image size increases from 49p to 2160p, the total energy consumption increases 45x when image size increases 44x times to process an image using an encoder with 49 comparators running at 1 GHz clock frequency.

As shown in \textit{Fig.\ref{fig:encoderResult_varying_imgSize_processingTime}}, when the image size increases from 49p to 2160p, there is 44x times increase in image size lead to 45x increase in processing time. 

\begin{figure}[h!]
    \begin{center}    \includegraphics[width=0.5\textwidth]{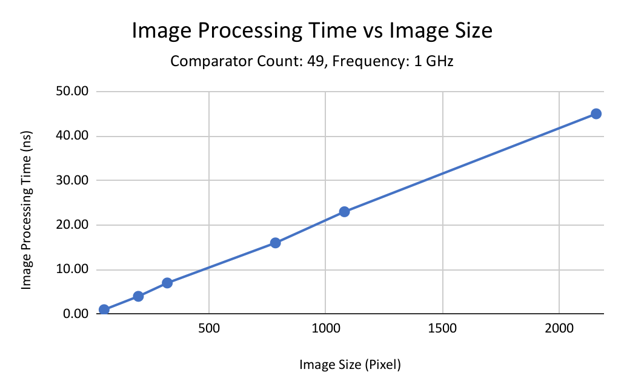}
    \end{center}
    \caption{Waveform analysis of total energy variation with respect to the varying input image size for 49 comparator count running at 1 GHz clock frequency.}    \label{fig:encoderResult_varying_imgSize_processingTime}
\end{figure}

\begin{figure}[h!]
    \begin{center}    \includegraphics[width=0.5\textwidth]{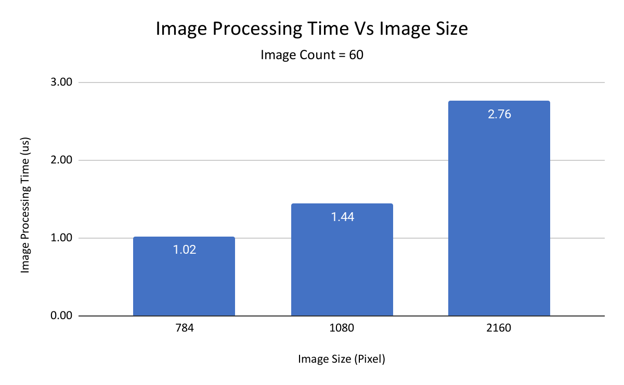}
    \end{center}
    \caption{Waveform analysis of processing time variation with respect to the varying input image size and image count of 60 for 49 comparator count running at 1 GHz clock frequency.}    \label{fig:encoderResult_imgSize_60_processingTime}
\end{figure}

We processed images having a refresh rate of 60Hz with 784p, 1080p, and 2160p image size processed with an encoder of 49 comparators running at 1GHz clock frequency to obtain a processing time of 60 images. Processing time for 60 images increases 2.7x with the image size increases 2.7x with processing time in the range of microseconds as shown in \textit{Fig.\ref{fig:encoderResult_imgSize_60_processingTime}}.

We obtain the image count to be processed in 1 ms processing time with 49 comparator encoders running at 1 GHz clock frequency, processing 784p, 1080p, and 2160p image size. Image count decreases 0.36x times as image size increases by 2.7x times as shown in \textit{Fig.\ref{fig:encoderResult_availProcessingTime_1ms_60_imgCount}}.

\begin{figure}[h!]
    \begin{center}    \includegraphics[width=0.5\textwidth]{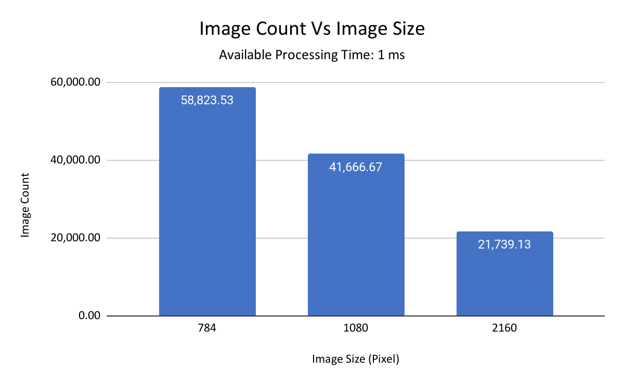}
    \end{center}
    \caption{Waveform analysis of image count variation with respect to the varying input image size and available processing time of 1 ms for 49 comparator count running at 1 GHz clock frequency.}    \label{fig:encoderResult_availProcessingTime_1ms_60_imgCount}
\end{figure}

\subsection{Gamma Cycle Relaxation Analysis}

\subsubsection{Functional Verification}
\hfill

The designs passed verification in both system-level and unit tests. All three test cases proved to properly generate or not generate the grst signal based on column outputs Fig. \ref{fig:verif}.

\begin{figure}[h!]
    \begin{center}    \includegraphics[width=0.5\textwidth]{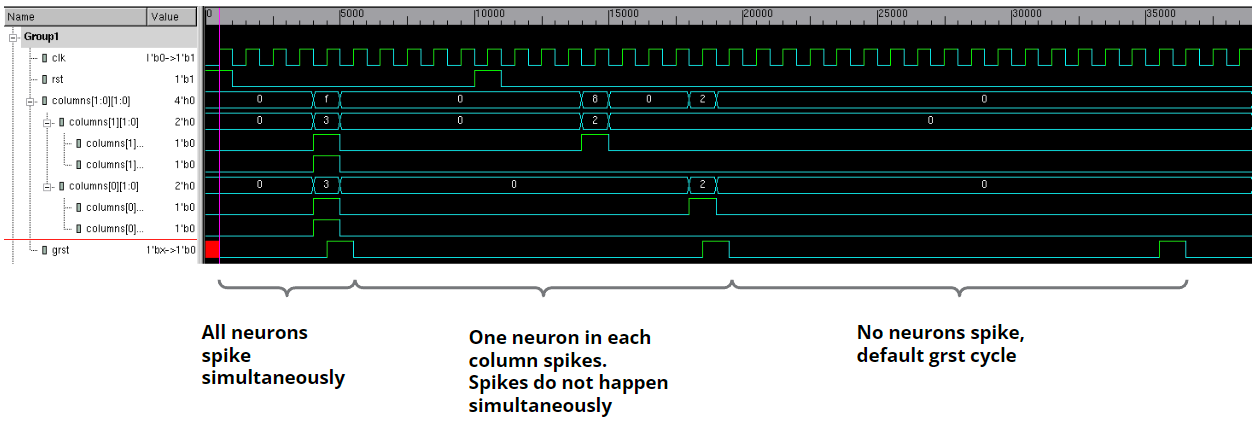}
    \end{center}
    \caption{Waveform analysis of the three performed system-level tests for the grst\_controller and grst\_generator modules.}    \label{fig:verif}
\end{figure}

\subsubsection{Design Costs}
\hfill

\textbf{Scaling with Neurons per column:} Plot in \textit {Fig.\ref{fig:neuron_col_energy}} shows, energy costs with scaling of the number of neurons per column in a fixed 676 column size. Energy consumption decreases with a decrease in the number of neurons but not linearly, at reduced neuron/column and the change in energy costs is minimal.

\begin{figure}[h!]
    \begin{center}    \includegraphics[width=0.5\textwidth]{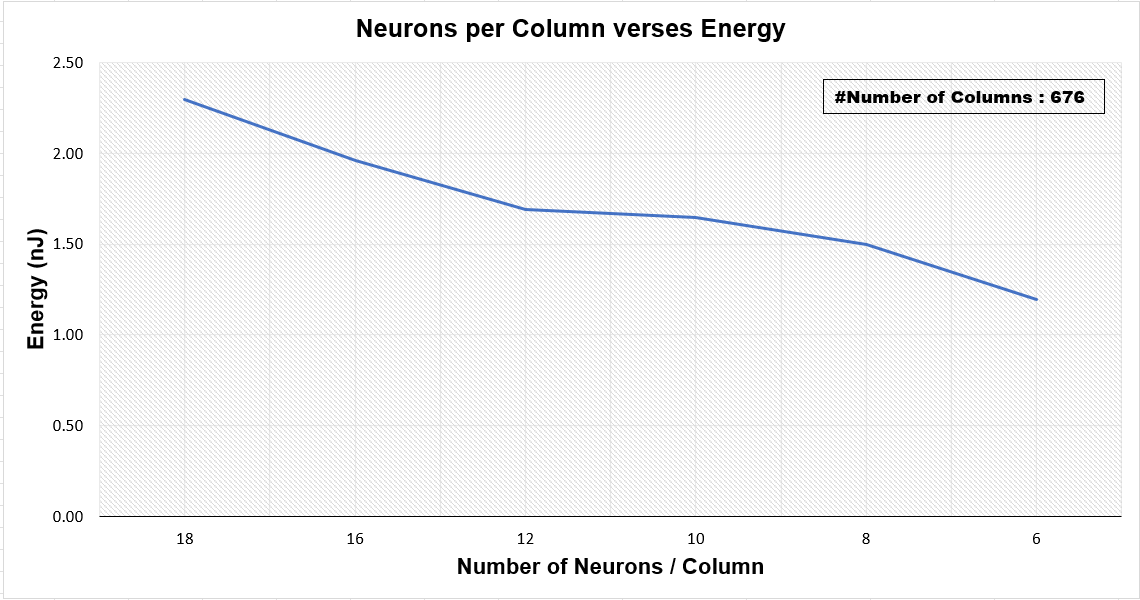}
    \end{center}
    \caption{neurons per column versus energy costs.}    \label{fig:neuron_col_energy}
\end{figure}

\textbf{Scaling with Column size:} In plot \textit{Fig.\ref{fig:num_col_energy}} shows energy costs with scaling of the number of columns in a fixed neuron size of 12. Energy consumption decreases with a decrease in the number of columns and decreases in energy are linear even at reduced neuron/column.
Critical path delay results in \textit{Fig.\ref{fig:num_col_critical_path}}, depicts the scaling of path delay with the number of columns. The result shows that path delay decreases linearly for column size decreased from $676$ to $484$, whereas delay remains nearly constant in further scaling down with gamma column size.

\begin{figure}[h!]
    \begin{center}    \includegraphics[width=0.5\textwidth]{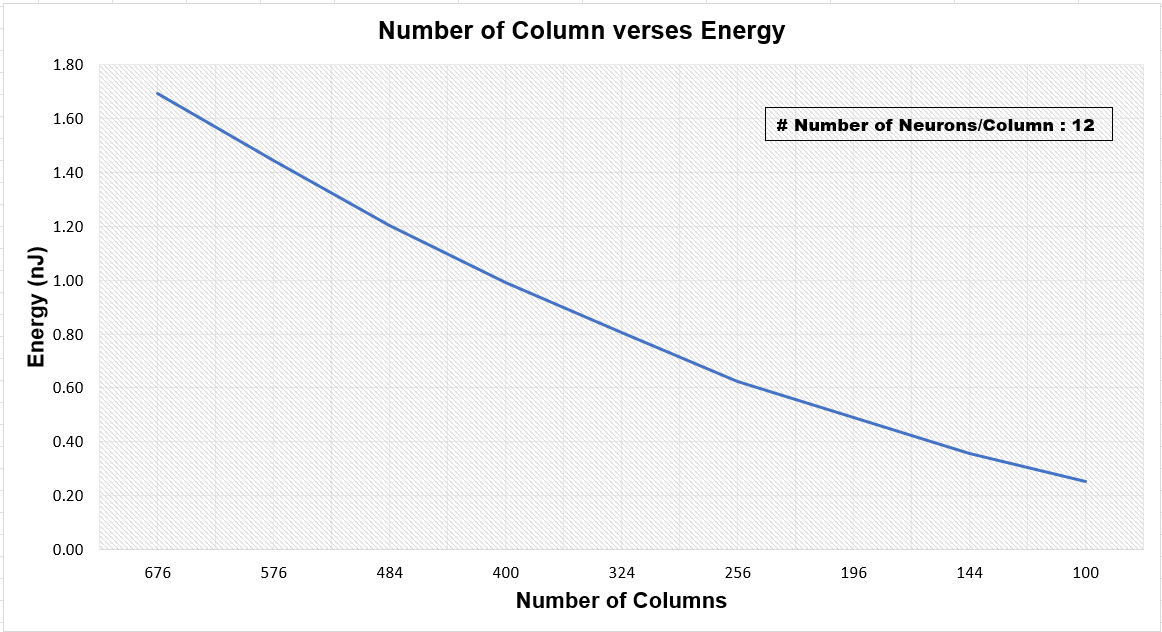}
    \end{center}
    \caption{number of columns versus energy costs.}    \label{fig:num_col_energy}
\end{figure}

\begin{figure}[h!]
    \begin{center}    \includegraphics[width=0.5\textwidth]{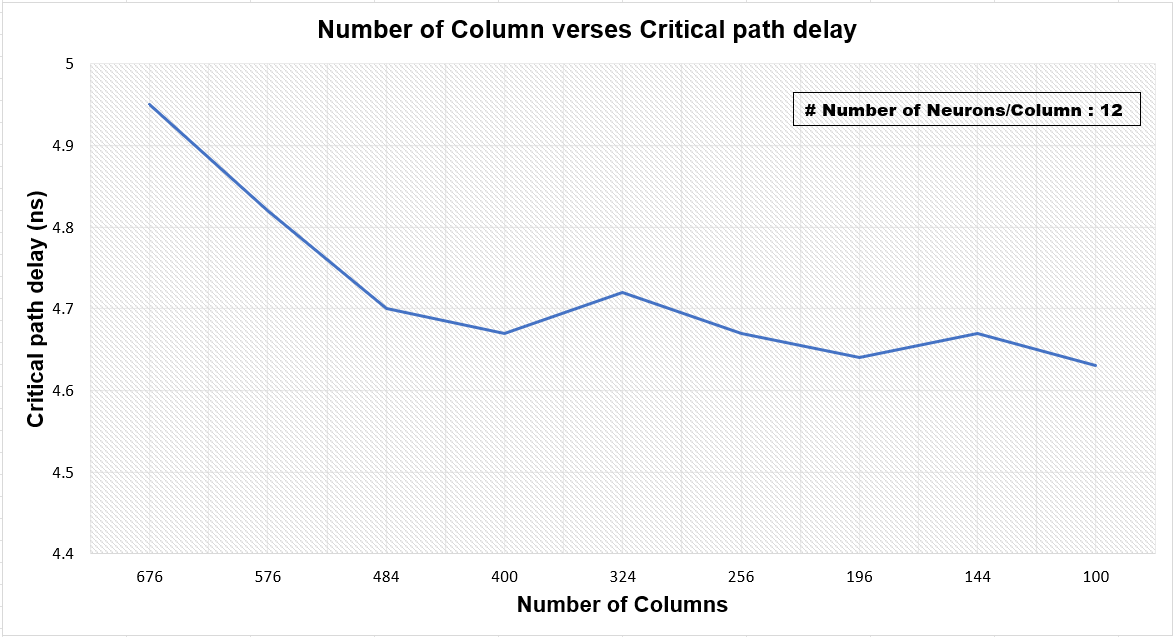}
    \end{center}
    \caption{Number of columns versus critical path delay.}    \label{fig:num_col_critical_path}
\end{figure}

\textbf{Scaling with Gamma controller size:} Gamma controller is characterized for a number of columns and number of neurons per column ranging from 676 columns x 12 neurons per column with different sizes of columns and neurons per column (1) larger 676x12 columns (2) medium 324x10 columns (3) small 121x6 columns. Design is evaluated for die area, energy costs, and path delay.
In \textit{Fig.\ref{fig:grst_size_energy}} shows that energy costs depict a declining trend with downscaling of gamma controller size from large, to medium and small, whereas \textit{Fig.\ref{fig:grst_size_critical_path}} plot shows that the minimum critical path delay is at medium gamma controller size dimension 324x10 column. From the perspective of energy costs and critical path delay trade-offs suggest that a medium-sized controller is optimal design.
\begin{figure}[h!]
    \begin{center}    \includegraphics[width=0.5\textwidth]{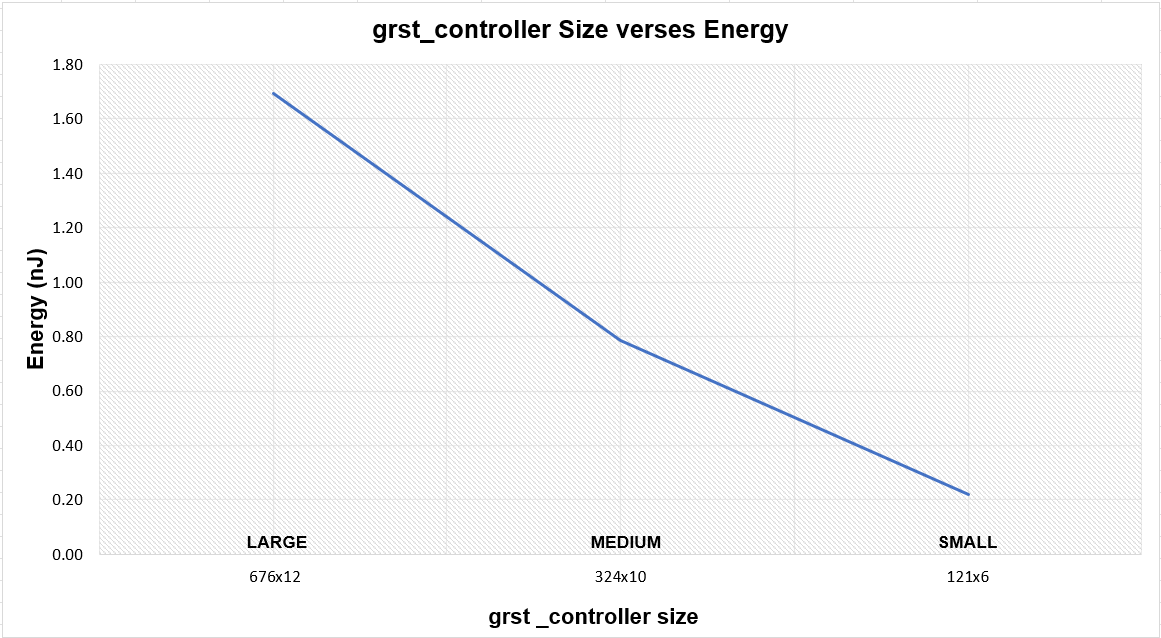}
    \end{center}
    \caption{gamma reset signal module size versus energy costs.}    \label{fig:grst_size_energy}
\end{figure}

\begin{figure}[h!]
    \begin{center}    \includegraphics[width=0.5\textwidth]{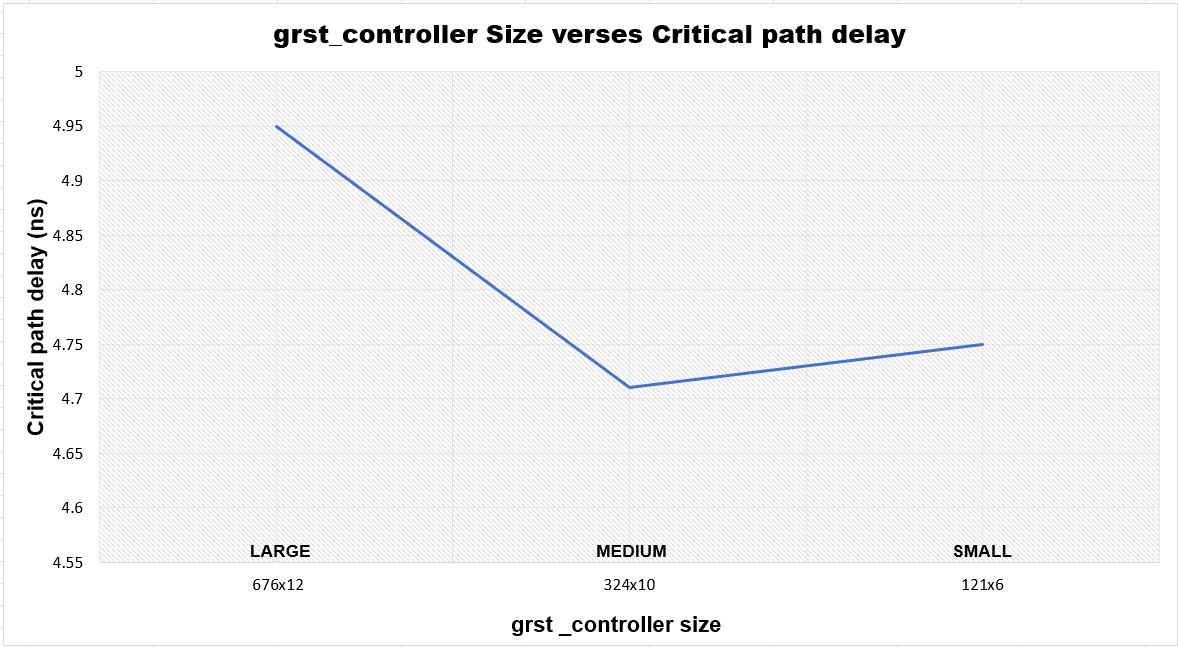}
    \end{center}
    \caption{gamma reset signal module size versus critical path delay.}    \label{fig:grst_size_critical_path}
\end{figure}


\subsubsection{Viability}
\hfill
 
The pattern seen for all RNL neural networks set to a threshold of 4000 was that of output spike times that start at infinity, and after some training time stabilize at around time 5 units as in \textit{Table \ref{tab3}}. After training, the test phase generated outputs at around the same time as the stabilized spike time from training time 5 units as in \textit{Table \ref{tab4}}.

\begin{table}[htbp]
\caption{Training - RNL Spiketime Occurrences With Threshold 4,000}
\begin{center}
\begin{tabular}{|c|c|c|c|c|}

\hline
\textbf{Spike Time} & \textbf{Linear} & \textbf{Log} & \textbf{PosNeg} \\

\hline
\text{5} & \text{7,869} & \text{8,058} & \text{8,063} \\
\hline
\text{6} & \text{1,279} & \text{32} & \text{27}\\
\hline
\text{7-17} & \text{7} & \text{0} & \text{0}\\
\hline
\text{inf} & \text{845}  & \text{1,910} & \text{1,910}\\
\hline
\textbf{Total} & \textbf{10,000} & \textbf{10,000} & \textbf{10,000} \\
\hline

\end{tabular}
\label{tab3}
\end{center}
\end{table}

\begin{table}[htbp]
\caption{Testing - RNL Spiketime Occurrences With Threshold 4,000}
\begin{center}
\begin{tabular}{|c|c|c|c|c|}

\hline
\textbf{Spike Time} & \textbf{Linear} & \textbf{Log} & \textbf{PosNeg} \\

\hline
\text{5} & \text{8,390} & \text{9,909} & \text{9,922} \\
\hline
\text{6} & \text{1,610} & \text{91} & \text{78}\\
\hline
\textbf{Total} & \textbf{10,000} & \textbf{10,000} & \textbf{10,000} \\
\hline

\end{tabular}
\label{tab4}
\end{center}
\end{table}

All RNL networks with a threshold of 400 produced output spikes at time 0 for all inputs over training and testing datasets.

In trained Log and PosNeg encoded networks, more than 99\% of spikes occurred at time 5 units, leading to a potential performance increase of 68\% (over the 16 clock cycles long default Gamma cycle) per Gamma cycle each Gamma cycle in a trained network \textit{Table \ref{tab5}}. The eventual spike time stabilization at a time less than the maximum gamma cycle time, demonstrates that relaxing gamma cycles has the potential to generate huge performance improvements on both training and trained networks.

\begin{table}[htbp]
\caption{Purity of Different Encoders}
\begin{center}
\begin{tabular}{|c|c|c|c|c|}

\hline
\textbf{Threshold} & \textbf{Linear} & \textbf{Log} & \textbf{PosNeg} \\

\hline
\text{400} & \text{0.468} & \text{0.6083} & \text{0.5951} \\
\hline
\text{4000} & \text{0.4444} & \text{0.5081} & \text{0.5123}\\
\hline

\end{tabular}
\label{tab5}
\end{center}
\end{table}

This interesting result showed that the Logarithmic encoding appeared to outperform the standard PosNeg encoding in purity at the original threshold of 400. This opens the possibility of worthwhile investigating optimal encoders for discussed RNL networks.

\section{Future Work}
An obvious next step for both systems would be to integrate them into the C3S environment. Performing an integration and tests for the encoder would potentially allow for future tests of the environment to be conducted on datasets such as MNIST in a simulated or FPGA-based environment, a significant next step in bringing the system in the direction of production. For the Gamma cycle controller and generator, though we have evidence to support the theory of the potential usefulness of the Gamma cycle relaxation, it remains to be proven with the STDP methodology used by C3S. Since the tested STDP performed with reference \cite{9516717}, is continuous and float-based, effects could be noticed as soon as weights passed a tipping point for training. However, for the lengthy, integer-based STDP updates, it may take longer to see benefits.

For the pos-neg encoder, the unit can provide better PPA results, with intelligent comparators capable of identifying and not processing undefined input, so as to provide reduced energy values independent of comparator count. 
It would also be interesting to explore different encodings and their effectiveness. Once an optimal encoding is found, the encoder module could be adapted to perform that encoding, and measurements for physical parameters could be re-taken.

\section{Conclusions}
Both the encoder and Gamma cycle control system have the potential to be useful additions to the C3S code base \cite{9516717}. 

Encoding from binary to spike times is an essential means for communication and data transmission between the two worlds of existing media formats and the evolving TNN infrastructure. This value has been recognized by groups such as BrainChip through their inclusion of such encoders on their novel Akida processor \cite{BrainChip}. A binary-to-spike encoder should be added to any upcoming neuromorphic system, for the world we live in is rife with data formats that do not fit well into a TNN style of data processing. 

The control and potential shortening of Gamma cycles possess the potential to take networks made with C3S columns and layers and improve the speed at which they perform their learning objectives. The potential for reducing the duration of Gamma cycles (and thus increase performance by) by upwards of 68\% percent, is significant. Additionally, the inclusion of such a control architecture may bring neuromorphic systems one step closer to the simulation of actual animal brains, which themselves do not always work on consistent frequencies.
The benefits of our work are products as modules that add to furthering of neuromorphic architectural research and should be integrated where possible to augment existing TNN neuromorphic systems.

\bibliography{references}

\begin{thebibliography}{1}

\bibitem{9516717}
Harideep Nair, John~Paul Shen, and James~E. Smith.
\newblock A microarchitecture implementation framework for online learning with
  temporal neural networks.
\newblock In {\em 2021 IEEE Computer Society Annual Symposium on VLSI
  (ISVLSI)}, pages 266--271, 2021.

\bibitem{article}
Scott Purdy.
\newblock Encoding data for htm systems.
\newblock 02 2016.

\bibitem{FRIES2007309}
Pascal Fries, Danko Nikolić, and Wolf Singer.
\newblock The gamma cycle.
\newblock {\em Trends in Neurosciences}, 30(7):309--316, 2007.
\newblock July INMED/TINS special issue—Physiogenic and pathogenic
  oscillations: the beauty and the beast.

\bibitem{Burwick2009GammaOA}
Thomas Burwick.
\newblock Gamma oscillations as integrators of local competition for activity
  and global competition for coherence.
\newblock {\em BMC Neuroscience}, 10:1--2, 2009.

\bibitem{Smith2017}
James~E. Smith.
\newblock {\em Biological Overview}, pages 47--77.
\newblock Springer International Publishing, Cham, 2017.

\bibitem{3}
James McCaffrey.
\newblock Preparing mnist image data text files.
\newblock
  \url{https://visualstudiomagazine.com/articles/2022/02/01/preparing-mnist-image-data-text-files.aspx},
  Jan 2022.

\bibitem{4}
Yann LeCun.
\newblock The mnist database of handwritten digits.
\newblock \url{http://yann.lecun.com/exdb/mnist/}.

\bibitem{BrainChip}
BrainChip.
\newblock Akida neural processor soc.
\newblock \url{https://brainchip.com/akida-neural-processor-soc/}, March 2023.

\end{thebibliography}
\bibliographystyle{unsrt}

\end{document}